# Efficient shortcut techniques in evanescently coupled waveguides

Koushik Paul and Amarendra K. Sarma*

Department of Physics, Indian Institute of Technology Guwahati, Guwahati, Assam, India

*Email: *aksarma@iitg.ernet.in*

**Abstract.** Shortcut to Adiabatic Passage (SHAPE) technique, in the context of coherent control of atomic systems has gained considerable attention in last few years. It is primarily because of its ability to manipulate population among the quantum states infinitely fast compared to the adiabatic processes. Two methods in this regard have been explored rigorously, namely the transitionless quantum driving and the Lewis-Reisenfeld invariant approach. We have applied these two methods to realize SHAPE in adiabatic waveguide coupler. Waveguide couplers are integral components of photonic circuits, primarily used as switching devices. Our study shows that with appropriate engineering of the coupling coefficient and propagation constants of the coupler it is possible to achieve efficient and complete power switching. We also observed that the coupler length could be reduced significantly without affecting the coupling efficiency of the system.

## 1. Introduction

In recent years the methods of shortcut to adiabatic passage (SHAPE) have gained considerable attention owing to its many possible applications in atomic and optical physics [1-8]. Shortcut methods originated due to the requirement of speeding up the adiabatic processes in atomic and quantum physics [9]. Transitionless quantum driving (TQD) and Lewis-Reisenfeld Invariant (LRI) methods are most widely used in this regard. The TQD approach drives a quantum system in a way such that adiabatic states become stationary enabling infinitely fast population transfer among the diabatic states [1,10]. On the other hand, the LRI approach exploits the property of dynamical invariants of the system to inverse engineer the conditions for instantaneous population exchange between the states. These methods have already been used in various contexts such as: creation of entanglement between atoms in cavity [11], atom cooling in harmonic traps [12] and control of spin in quantum dot [13], to mention a few. It is interesting to note that, drawing inspiration from many analogies with quantum physics, these methods have been applied even in wave-optics [14-16]. In this regard, coupled waveguides in integrated optics are particularly interesting due to its tremendous practical applications [17,18]. In general, the function of a waveguide coupler is to split coherently an optical field incident on one of the input ports and direct the two parts to the output ports. Adiabatic passage technique has been exploited successfully in two and three waveguide couplers in order to study the eigenmode evolution of optical power [19-23]. For a sufficiently long coupler, where adiabaticity is satisfied, the system follows its initial eigenmode, causing power transfer from one waveguide to the other. Large device length causes higher transmission loss and makes designing practical devices difficult. However, there are significant opportunities to make couplers more efficient and small in dimension using shortcut methods. Several new studies in this regard have been reported recently [24,25].

In this work we choose a waveguide coupler of length **L** consists of two single mode waveguides with **z** being the direction of propagation. The schematic of the system is shown in Fig.1. Waveguides are tapered in nature so that propagation constants $\boldsymbol{\beta_1}$ and $\boldsymbol{\beta_2}$ varies along **z**. Separation between the waveguides is not constant and the evanescent coupling coefficient between the waveguides ($\boldsymbol{\kappa}$) also varies along **z** and is maximum at length $\boldsymbol{L/2}$ where the separation is minimal.

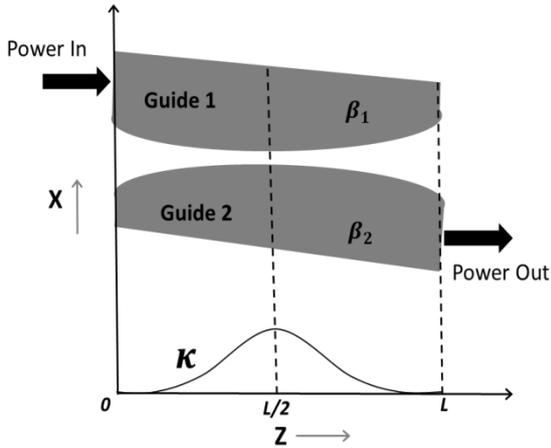

**Figure 1.** Schematic diagram for tapered waveguide coupler of length $L$. Propagation direction is along $z$ direction. $\beta_1$ and $\beta_2$ are the propagation constants. Separation between the waveguides varies along $z$ and the coupling coefficient $\kappa$ attains a maximum at $z = L/2$.

## 2. Shortcut Methods

To illustrate the basic principles, let us start from the coupled mode theory for coupled optical waveguides. Coupled mode equations are expressed as $d/dz[a_1(z), a_2(z)]^T = -i\, H(z)[a_1(z), a_2(z)]^T$. We recognize $H(z)$ as the Hamiltonian and is given by:

$$H(z) = \begin{pmatrix} \Delta(z) & \kappa(z) \\ \kappa(z) & -\Delta(z) \end{pmatrix} \quad (1)$$

Here $a_1$ and $a_2$ are the modal amplitudes of the respective waveguides. $\Delta$ is the mismatch parameter, defined as $\Delta = (\beta_2 - \beta_1)/2$ and $z$ is the direction of propagation. When the rate of variation of $\Delta(z)$ and $\kappa(z)$ is small enough with respect to $z$, the system undergoes adiabatic evolution along $z$. Adiabatic condition for such system is $\kappa_0 L \gg 1$, where $\kappa_0$ is the coupling coefficient at $L/2$. This clearly suggests the requirement for large length of the coupler. By using shortcut methods, we will show how to circumvent this issue, even when adiabatic condition is violated.

*2.1 Lewis-Reisenfeld Invariant approach*
The Hamiltonian (1) can be written in the following form:

$$H(z) = \kappa(z)\sigma_x + 0\,\sigma_y + \Delta(z)\sigma_z \quad (2)$$

Here $\sigma_x$, $\sigma_y$ and $\sigma_z$ are the well-known Pauli matrices for spin $\frac{1}{2}$ particles. These operators satisfy Lie algebra: $[\sigma_i, \sigma_j] = 2i\epsilon_{ijk}\sigma_k$. The Hamiltonian satisfies SU (2) symmetry and hence there exist an invariant which would satisfy the usual invariant equation: $dI(z)/dz = \partial I(z)/\partial z - i[I(z), H(z)]$. According to Lewis-Reisenfeld theory [26], this invariant can be written as follows:

$$I(z) = \frac{\Omega}{2} \begin{pmatrix} \cos \gamma(z) & \sin \gamma(z) \, e^{-i\beta(z)} \\ \sin \gamma(z) \, e^{i\beta(z)} & -\cos \gamma(z) \end{pmatrix} \tag{3}$$

Here $\Omega$ is an arbitrary constant which has the dimension of $\kappa(z)$. $\gamma(z)$ and $\beta(z)$ are the parameters which characterizes $I(z)$ and satisfies the following conditions:

$$\dot{\gamma}(z) = 2\kappa(z) \sin \beta(z) \tag{4a}$$
$$\left(2\Delta(z) + \dot{\beta}(z)\right) \sin \gamma(z) = 2\kappa(z) \cos \gamma(z) \cos \beta(z) \tag{4b}$$

where overdot represents derivative with respect to $z$. It is important to note that $I(z)$ and $H(z)$ does not commute normally. To make $I(z)$ and $H(z)$ commute at the boundaries so that the eigenstates exactly match at the ends of the coupler, we impose $[H(0), I(0)] = 0$ and $[H(L), I(L)] = 0$. With straightforward calculations we obtain the following constraints:

$$[\kappa(z) \sin \gamma(z) \sin(z)]_{z=0,L} = 0 \tag{5a}$$
$$\left[\sqrt{2}\kappa(z) \cos \gamma(z) - \Delta(z) \sin \gamma(z) \, e^{\pm i\beta(z)}\right]_{z=0,L} = 0 \tag{5b}$$

These constraints help us to determine the required boundary conditions for $\beta(z), \gamma(z)$ and $\kappa(z)$. We find that, the boundary conditions can be satisfied only if $\kappa(0) = \kappa(L) = 0$ and $\sin \gamma(0) = \sin \gamma(L) = 0$. $\beta(z)$ helps us to configure $\kappa(z)$ and $\Delta(z)$. It is to be noted that $\beta$ cannot be chosen to be zero as $\kappa(z)$ is finite. We chose $\beta(z)$ such that $\kappa(z)$ keeps its amplitude minimal. With the above considerations, we set the boundary conditions as follows:

$$\gamma(0) = \pi; \quad \gamma(L) = 0; \quad \dot{\gamma}(0) = 0; \quad \dot{\gamma}(L) = 0 \tag{6}$$
$$\beta(0) = -\frac{\pi}{2}; \quad \beta(L) = -\frac{\pi}{2}; \quad \dot{\beta}(0) = \frac{3\pi}{2L}; \quad \dot{\beta}(L) = -\frac{3\pi}{2L} \tag{7}$$

Using the above boundary conditions one can easily construct the parameters $\gamma$ and $\beta$, and thereby all the other necessary parameters to study the evolution of optical power within the waveguides.

### 2.2 Transitionless Driving algorithm

According to Berry's Transitionless driving algorithm [10], when the adiabatic conditions are not satisfied (for relatively small lengths), we need to find an additional interaction term in order to avoid non-adiabatic effects. To construct such interaction term we use unitary transformation to go to the adiabatic basis, $\{A_j\}$, using: $[A_1, A_2]^T = U_0^{-1} [a_1, a_2]^T$. The driving Hamiltonian is expressed as: $H_a = i \sum_j |\partial_z A_j\rangle\langle A_j|$, which when written in explicit form looks like:

$$H_a(z) = \begin{pmatrix} 0 & -i\dot{\theta}/2 \\ i\dot{\theta}/2 & 0 \end{pmatrix} \tag{8}$$

Here $\theta(z) = \tan^{-1} \kappa(z)/\Delta(z)$ is the angle of mixing. $\dot{\theta}$ represents additional coupling between those waveguides. We add $H_a(z)$ back to the original Hamiltonian to find an effective coupling that can be realized through appropriate engineering of the coupler. Using the unitary transformation, $U_1 = e^{-i\phi/2}|a_1\rangle\langle a_1| + e^{i\phi/2}|a_2\rangle\langle a_2|$ the effective Hamiltonian can be expressed as follows:

$$H_{eff}(z) = \begin{pmatrix} \Delta_{eff}(z) & \kappa_{eff}(z) \\ \kappa_{eff}(z) & -\Delta_{eff}(z) \end{pmatrix} \tag{9}$$

Here $\kappa_{eff} = \sqrt{\kappa^2 + \dot{\theta}^2/4}$ and $\Delta_{eff} = \Delta - \dot{\phi}/2$. The unitary transformations $U_0$ and $U_1$ are connected via the parameters $\theta$ and $\phi$. These parameters need to be adjusted such that the adiabatic bases of the unitary transformations are equivalent at the boundary. This leads to the following boundary conditions: $\dot{\theta}(0) = \dot{\theta}(L) = 0$ and $\theta(0) = \theta(L) = 2\pi$. However in our case $\phi$ can be chosen arbitrarily as it has no effect on power evolution.

## 3. Results and Discussion

Using the conditions in Eq. (6) and (7), $\gamma(z)$ and $\beta(z)$ can be constructed. We followed the polynomial ansatz to interpolate them at the intermediate points. We choose them as $\gamma(z) = \sum_{i=0}^{3} g_i z^j$ and $\beta(z) = \sum_{i=0}^{3} b_i z^j$, where $g_j$s and $b_j$s are determined using boundary conditions.

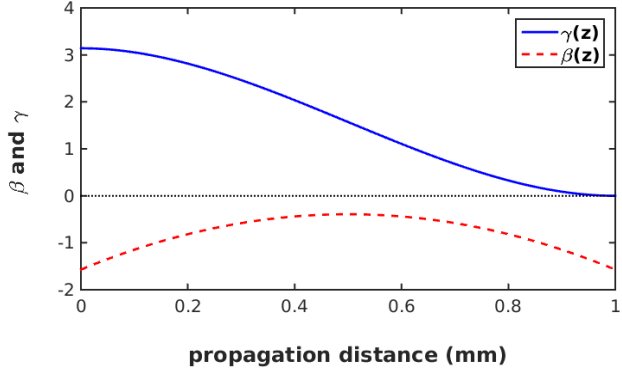

**Figure 2.** (color online) profile of $\beta(z)$ (dashed) and $\gamma(z)$ (solid) determined through polynomial ansatz with the coefficients determined using the boundary conditions.

Fig. (2) shows the spatial profile of $\gamma$ and $\beta$. The mismatch parameter $\Delta$ and coupling coefficient , as determined through the invariant method, are shown in Fig. (3a). The profile of $\Delta_{\text{eff}}$ and $\kappa_{eff}$, determined from the transitionless driving algorithm is shown in Fig. (3b), where $\Delta$ and $\kappa$ are chosen as per the famous Allen-Eberly scheme [27]:

$$\kappa(z) = \kappa_0 \text{sech}[(2\pi(z - z_0))/L], \qquad \Delta(z) = \Delta_0 \tanh[2\pi(z - z_0)/L] \qquad (10)$$

We have taken: $\kappa_0, \Delta_0 = 1\ mm^{-1}$ and $z_0 = L/2$. Similarity between both the methods are evident from Fig. (3a) and (3b). The strength of the couplings determined from both the methods is almost same, however the mismatch is much larger in case of the invariant method.

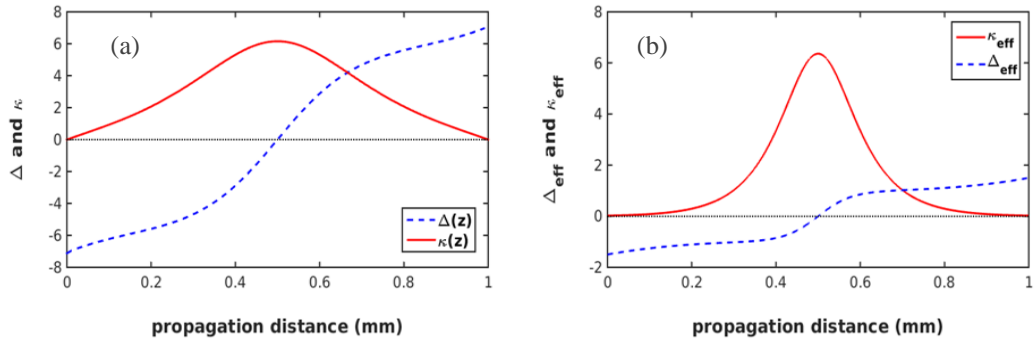

**Figure 3.** (color online). Spatial profile of coupling and mismatch. (a) $\Delta(z)$ (dashed) and $\kappa(z)$ (solid) determined through Invariant based method, (b) $\Delta_{eff}(z)$ (dashed) and $\kappa_{eff}(z)$ (solid) determined from transitionless driving method.

We have numerically solved the Master equation for density matrix to study the power evolution inside the waveguides [24]. Fig. (4) shows fractional power evolution, defined as $P_2(z)/P_1(0)$, within the coupler. Optical power is launched in the first waveguide while the other one is kept empty. For adiabatic regime, in Fig. (4a) we have chosen a 100 mm coupler device, which is sufficiently large so that the adiabatic condition is being satisfied. Complete power transfer is observed. Fig. (4b) shows the power evolution through the transitionless driving method which follows the adiabatic path exactly but within a coupler size of merely 1mm.

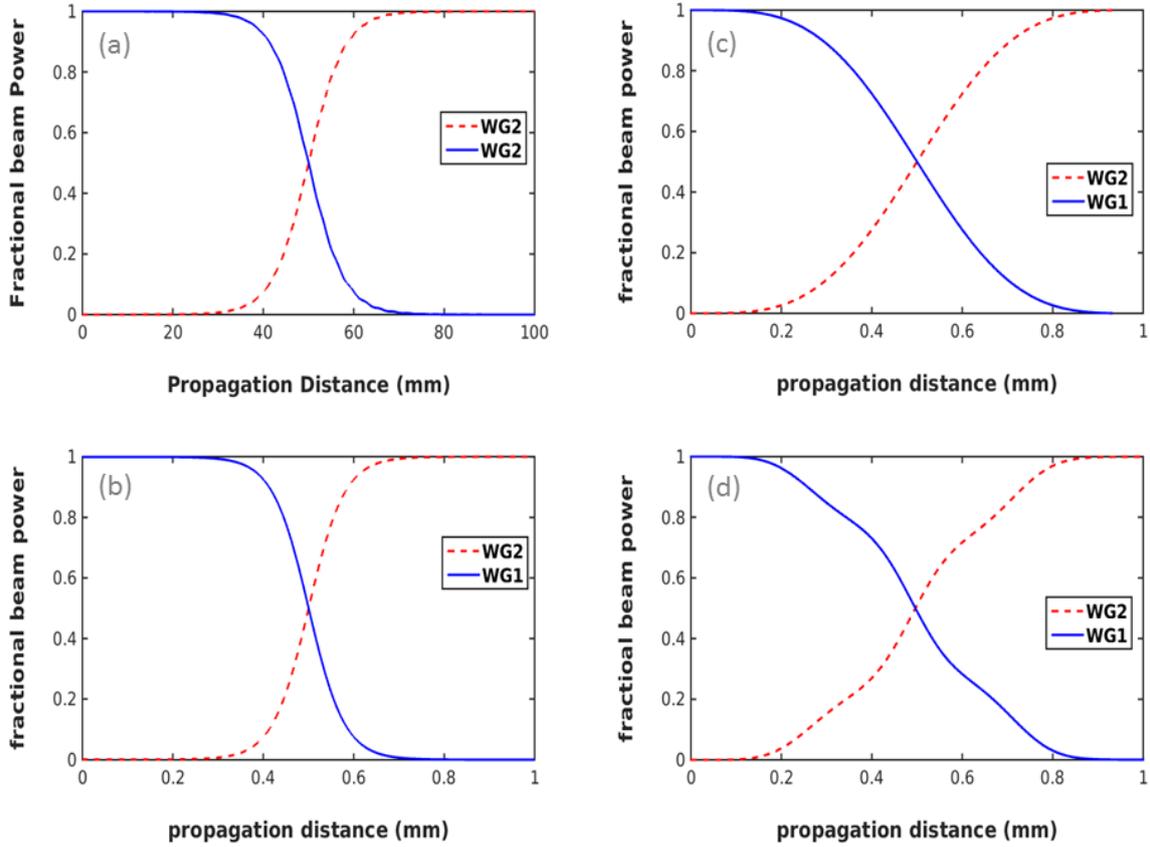

**Figure 4.** (color online) Spatial evolution of fractional beam power with respect to $z$ for (a) adiabatic case with $L = 100mm$, (b) Transitionless driving where $L = 1mm$, (c) & (d) Invariant based approach using the Hamiltonian and the invariant respectively, $L = 1mm$.

In Fig. (4c), $\kappa(z)$ and $\Delta(z)$ are taken according to Eq. (4) and used in the original adiabatic Hamiltonian. It exactly shows the adiabatic nature of the evolution but again only 1mm long coupler is enough to complete the power transfer. In Fig. (4d), we have used the Lewis-Reisenfeld invariant instead of the actual Hamiltonian, which precisely matches with Fig. (4c) at the boundaries but does not follow the adiabatic path.

## 4. Conclusions
In conclusion we have described two very efficient methods for power transfer in waveguide couplers using the analogy between quantum formalism and coupled mode theory for waveguides. We have shown that with judicious choice of coupling coefficient and mismatch profile it is possible to construct a coupler, which is considerably small in size compared to adiabatic couplers. With recent development of fabrication techniques, separation between the waveguides and profile of taper can be controlled accurately and hence designing $\kappa$ and $\Delta$, as predicted by these theories, is achievable. The proposed methods may be used to design couplers suitable for photonic circuits, of which couplers are integral components.

**Acknowledgement**

K. Paul would like to thank MHRD, Government of India for a research fellowship.